\documentclass[pra,showpacs,twocolumn]{revtex4}
\usepackage{bm,graphicx,amsmath}
\usepackage{amsfonts}
\usepackage{amssymb}
\usepackage{amsthm}
\begin{document}
\title{Jahn-Teller induced Berry phase in spin-orbit coupled
Bose-Einstein condensates}
\author{Jonas Larson$^1$ and Erik Sj\"oqvist$^2$}
\affiliation{$^1$NORDITA, 106 91 Stockholm, Sweden \\
$^2$Department of Quantum Chemistry, Uppsala University,
Box 518, Se-751 20 Uppsala, Sweden}
\date{\today}
\begin{abstract}
We demonstrate that Berry phases may greatly affect the dynamics
of spin-orbit coupled Bose-Einstein condensates. The effective model
Hamiltonian under consideration is shown to be equivalent to the
$E\times\varepsilon$ Jahn-Teller model first introduced in molecular
physics. The corresponding conical intersection is identified and the
Berry phase acquired for a wave packet encircling the intersection
studied. It is found that this phase manifests itself in the density
profile of the condensate, making it a directly measurable quantity
via time-of-flight detection. Moreover, the non-Abelian gauge
structure of the system is addressed and we verify how it affects
the dynamics.
\end{abstract}
\maketitle

\section{Introduction}
The realization of Bose-Einstein condensates \cite{bec} opened a new
door for the study of quantum phenomena. Since then the field has
seen a tremendous progress in terms of trapping, cooling and
manipulating atoms, covering cold atoms in optical lattices
\cite{maciek}, BCS-BEC crossover \cite{bcs} and multi-component
Bose-Einstein condensates \cite{spinorbec}. Moreover, it has been
demonstrated both theoretically \cite{ohberg1,gaugegen} and
experimentally \cite{gaugeexp} that ultracold multi-level atoms
moving in spatially dependent laser fields give rise to effective
gauge potentials. Considering four-level atoms in a tripod setup,
non-Abelian gauge fields are achieved \cite{gaugeref2}. The idea of
utilizing a four-level atom with three degenerate ground states
coupled to one excited state to generate non-Abelian gauge
potentials dates back to the work of Unanyan and co-workers
\cite{fourlevel}, where however the effective magnetic fields arise
from time-dependence of the Hamiltonian rather than spatial
dependence. The particular system configuration of
Ref.~\cite{gaugeref2} has turned out to be extremely rich, and a
series of papers investigating various aspects of the model have
been published recently. Among these are, spin-Hall effects
\cite{spinhall}, the Aharonov-Bohm effect \cite{berryCA1},
relativistic characteristics \cite{rev1,rev2}, spin-echo phenomena of
trapped fermions \cite{SO1}, spin dynamics \cite{spin}, expansion to
time-dependent laser fields \cite{timelaser}, novel phases of the
condensate \cite{SO2}, the structure of the energy spectrum
\cite{landau}, and effective magnetic monopoles \cite{mono}.

For certain laser arrangements, the energy dispersions possess a
point-degeneracy also termed {\it conical intersection} (CI)
\cite{baer1}. It is well known \cite{ABmol,gaugeref} that apart from
the dynamical phase, encircling a CI renders a Berry phase of
the state vector \cite{berry}. Differing from typical situations in
molecular and chemical physics, here the CI occurs in momentum and not
in position space. As a consequence, the Berry phase is more
directly manifested in the momentum wave function. Similar situations
appear in graphene \cite{graphene} and cold atoms in elaborate optical
lattice configurations \cite{cilattice}.

In the current work, we consider a harmonically trapped spinor
Bose-Einstein condensate in the presence of spatially dependent laser
fields such that a CI is recovered in momentum space. By taking the
harmonic confinement of the condensate into account, the resulting
Hamiltonian is equivalent to the $E\times\varepsilon$ Jahn-Teller one
\cite{lh}, but with position and momentum interchanged. Moreover, the
non-linear terms arising from atom-atom scattering are included. Such
terms are absent in for example CI-models in molecular physics
\cite{baer1,CIrev,molberry}.  Utilizing numerical wave packet
simulations, we thoroughly discuss the Berry phase and how it affects
the system dynamics, both on a short and long time scale. It is found
that over longer time periods, the population of the two phonon modes
is swapped, and in particular, the characteristics of this exchange
mechanism depend on the Berry phase. Various studies have been
concerned about Berry phases in Bose-Einstein condensates influenced
by slowly varying external fields
\cite{berrybec1,berrybec2}. Contrary to these references, in the
present model the Berry phase shows up in the internal spinor
dynamics as an effective spatially dependent gauge field, in analogy
with the Born-Oppenheimer scenario in molecular theory
\cite{ABmol}. We note that Berry phases were as well considered
in Refs.~\cite{berryCA1,rev2}, but without taking into account for a
trapping potential nor atom-atom collisions. A harmonic trapping
potential causes the momentum wave packet to spread, which has
important consequences for its dynamics.  In addition to investigating
the consequences of a non-zero Berry phase, we also study the
non-Abelian structure of the system. We show that one round-trip
of the CI clockwise or anti-clockwise brings about different final
states.

The paper is outlined as follows. In the next section, the model
system is introduced and the effective Gross-Pitaevskii equation
given. The equivalence with the $E\times\varepsilon$ Jahn-Teller
Hamiltonian is demonstrated and we discuss the form of the momentum
potential energy surfaces. Section \ref{sec3} is devoted to numerical
simulations. First, in Sec.~\ref{ssec3a}, the short time evolution is
considered, the non-cyclic Berry phase \cite{samuel88} for the spinor
state is computed, the impact of the Berry phase on the wave packet
dynamics is clarified, and we show how a time-of-flight measurement of
the condensate would reveal the presence of a Berry phase. The
following Sec.~\ref{ssec3b} deals with the long time scale
properties. A collapse-revival as well as a swapping phenomena is
found. The underlying non-Abelian structure is envisaged in
Sec.~\ref{ssec3c}. Last we finish with concluding remarks in
Sec.~\ref{sec4}.

\section{Model system}\label{sec2}
We begin by deriving the effective single atom Hamiltonian, and then
turn to the many-body case by including atom-atom scattering to obtain
a mean field Gross-Pitaevskii equation.

\begin{figure}
\centerline{\includegraphics[width=8cm]{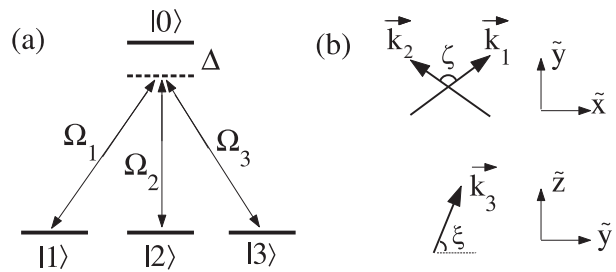}} \caption{Atomic
(a) and laser (b) configurations.} \label{fig1}
\end{figure}

The setup is detailed in Fig.~\ref{fig1} (a). Three degenerate meta
stable Zeeman ground states are dipole coupled to an excited state via
three respective external lasers. Using the notations of the figure
and imposing the rotating wave approximation, the interaction
Hamiltonian reads
\begin{eqnarray}\label{iham}
H_{\mathrm{I}} & = &
\hbar\Delta|0\rangle\langle0| +
\hbar \big( \Omega_1|0\rangle\langle1| + \Omega_2 |0\rangle\langle2|
\nonumber \\
 & & + \Omega_3|0\rangle\langle3| + \textrm{h.c.} \big) ,
\end{eqnarray}
which defines the full Hamiltonian $H = \tilde{\vec{p}}\,^2/(2m) +
H_I$, $\tilde{\vec{p}}$ and $m$ being the momentum and mass,
respectively, of the atom. The coupling strengths, being
proportional to the laser amplitudes, are parameterized as
$\Omega_1=\Omega\sin\theta\cos\varphi e^{iS_1}$,
$\Omega_2=\Omega\sin\theta\sin\varphi e^{iS_2}$ and
$\Omega_3=\Omega\cos\theta e^{iS_3}$ with
$\Omega=\sqrt{|\Omega_1|^2+|\Omega_2|^2+|\Omega_3|^2}$. Here, $\theta$,
$\varphi$ and $S_i$ are allowed to be spatially dependent. The
interaction part $H_{\mathrm{I}}$ of the full Hamiltonian may be
readily diagonalized, rendering the eigenstates $|w_0\rangle$,
$|w_3\rangle$, $|u_1\rangle$, and $|u_2\rangle$, where in particular
the two states
\begin{equation}
\begin{array}{lll}
|u_1\rangle & = &
\displaystyle{\frac{1}{\sqrt{2}}}\Big(\left[\sin\varphi
e^{i\pi/4} + \cos\!\theta\cos\!\varphi
e^{-i\pi/4}\right]e^{-iS_{13}}|1\rangle\\ \\ & & -\left[\cos\varphi
e^{i\pi/4}-\cos\!\theta\sin\!\varphi
e^{-i\pi/4}\right]e^{-iS_{23}}|2\rangle \\ \\ & & -\sin\!\theta
e^{-i\pi/4}|3\rangle\Big), \\ \\
|u_2\rangle & = &
\displaystyle{\frac{1}{\sqrt{2}}}\Big(\left[\cos\!\theta\cos\!\varphi
e^{-i\pi/4}-\sin\varphi e^{i\pi/4}\right]e^{-iS_{13}}|1\rangle\\ \\ &
& +\left[ \cos\varphi e^{i\pi/4}+\cos\!\theta\sin\!\varphi
e^{-i\pi/4}\right]e^{-iS_{23}}|2\rangle \\ \\ & & -\sin\!\theta
e^{-i\pi/4}|3\rangle\Big), \\ \\
\end{array}
\end{equation}
are degenerate with zero eigenvalue, and here $S_{ij}=S_i-S_j$.
These two states are termed {\it dark states} as they do not couple
to the excited state $|0\rangle$ via $H_{\mathrm{I}}$. Due to the
shifted energy of the remaining two states $|w_0\rangle$ and
$|w_3\rangle$ compared to $|u_1\rangle$ and $|u_2\rangle$, they can
be adiabatically eliminated from the Hamiltonian provided that
$\Delta\gg\Omega$, ending up with an effective two-level problem for
$|u_1\rangle$ and $|u_2\rangle$.

Before writing down the resulting Hamiltonian we specify the laser
configurations \cite{SO1}, see Fig.~\ref{fig1} (b). First we assume
$S_1=S_2$, $S_{31}=mv_s\tilde{y}/\hbar$ and $\varphi=mv_\varphi
\tilde{x}/\hbar$ and $\theta\in[0,\pi]$ some constant. Two
lasers, which govern the coupling of $|1\rangle$ and $|2\rangle$ to
$|0\rangle$, are propagating along $\vec{k}_1$ and $\vec{k}_2$ in
the $\tilde{x}\tilde{y}$-plane, where the angle of intersection
between them is $\zeta$. In the $\tilde{x}$-direction, these two
lasers are standing waves, however mutually shifted in phase by
$\pi/2$, and propagating waves in the $\tilde{y}$-direction. A third
laser is a propagating wave along $\vec{k}_3$ in the
$\tilde{y}\tilde{z}$-plane, whose direction is determined by the
angle $\xi$ relative the $\tilde{y}$-axis. In terms of the angles
$\zeta$ and $\xi$ and the wave numbers $k_i = |\vec{k}_i|$ we have
\begin{equation}
\begin{array}{l}
S_1=S_2=k_1\tilde{y}\cos (\zeta/2), \\ \\
S_3=k_3\tilde{y}\cos\xi , \\ \\
\varphi=2k_1\tilde{x}\sin (\zeta/2).
\end{array}
\end{equation}

After the adiabatic elimination, the effective single particle
two-level Hamiltonian becomes \cite{SO1}
\begin{equation}\label{ham0}
H_{\mathrm{eff}} = \frac{\tilde{p}^2}{2m} +
\delta_0\hat{\sigma}_y-\tilde{v}_0\hat{\sigma}_x\tilde{p}_{\tilde{x}}-
\tilde{v}_1\hat{\sigma}_y\tilde{p}_{\tilde{y}},
\end{equation}
where $\delta_0$ is an effective Zeeman splitting,
$\tilde{v}_0=v_\varphi\cos\theta$, $\tilde{v}_1=v_s\sin^2\theta/2$
and $\hat{\sigma}_i$, with $i=x,y,z$ are the Pauli operators
$\hat{\sigma}_x = |u_1\rangle \langle u_2| + |u_2\rangle \langle
u_1|$, $\hat{\sigma}_y = -i|u_1\rangle \langle u_2| + i|u_2\rangle
\langle u_1|$, and $\hat{\sigma}_z = |u_1\rangle \langle u_1| -
|u_2\rangle \langle u_2|$. The $\tilde{p}$-dependence of the last
two terms originates from non-adiabatic couplings \cite{jonas1}, and
we note that these do not include $\tilde{p}_{\tilde{z}}$ and we may
therefore factor out the $\tilde{z}$-dependence. The spin-orbit
coupling is on a {\it Rashba}-form \cite{rashba1}, which has been
frequently studied in the context of semiconductor quantum-dots
\cite{rashba2,rashba3}.

We will further add a harmonic trapping potential (associated
with angular frequency $\omega$) for the atoms. This
was in fact also done in Ref.~\cite{rashba3}. However, for the
condensate considered in this work, such a trapping potential is
almost exclusively present in experiments, while this is not the case
for the electrons in semiconducting quantum-dots of
Ref.~\cite{rashba3}. By properly Stark shifting the internal atomic
states, individual trapping potentials can be achieved. In particular,
for certain constant Stark shifts, the Zeeman splitting term in the
Hamiltonian \cite{berryCA1,SO1} can be made to vanish, which we will
assume in the following.  It should be mentioned that in order to
justify adiabatic elimination of the two dark states, once the
harmonic trapping potential has been included, one should assume not
only $\Delta\gg\Omega$, but also $\Omega\gg\omega$ in order to ensure
the validity of the Born-Oppenheimer separation.

Imposing $s$-wave scattering between the atoms in terms of the
regular non-linear term gives a Gross-Pitaevskii equation. Defining
the spinor wave function
\begin{equation}
\tilde{\Psi}(\tilde{x},\tilde{y},t) \! = \!
\left[\!\!\begin{array}{c}\langle \tilde{x},\tilde{y}|
\langle u_1|\tilde{\Psi}(t)\rangle\\
\langle \tilde{x},\tilde{y}|\langle
u_2|\tilde{\Psi}(t)\rangle\end{array}\!\!\right]\!\! = \!
\left[\!\!\begin{array}{c}\tilde{\psi}_1(\tilde{x},\tilde{y},t)\\
\tilde{\psi}_2(\tilde{x},\tilde{y},t)\end{array}\!\!\right],
\end{equation}
where
$\psi_i(\tilde{x},\tilde{y},t)=\langle\tilde{x},\tilde{y}|\psi_i(t)\rangle$,
$i=1,\,2$ and the normalization reads $\int
d\tilde{x}d\tilde{y}|\Psi(\tilde{x},\tilde{y},t)|^2=1$, the
Gross-Pitaevskii equation takes the form
\begin{equation}\label{ham}
\begin{array}{lll}
i \hbar \frac{\partial}{\partial t}\tilde{\Psi}(\tilde{x},\tilde{y},t)
\!\!\! & = \!\!\!\!& \Big[-\frac{\hbar^2}{2m}
\left(\frac{\partial^2}{\partial \tilde{x}^2} +
\frac{\partial^2}{\partial \tilde{y}^2}\right) +
\frac{m\omega^2}{2}\left(\tilde{x}^2+\tilde{y}^2\right) \\ \\
& & \!\!\!\!-i\hbar \tilde{v}_0\boldsymbol{\sigma}_x
\frac{\partial}{\partial\tilde{x}} -i\hbar \tilde{v}_1
\boldsymbol{\sigma}_y \frac{\partial}{\partial\tilde{y}} \\ \\
& & \!\!\!\! +
\tilde{\Psi}^\dagger(\tilde{x},\tilde{y},t)
\mathbf{g} \tilde{\Psi}(\tilde{x},\tilde{y},t)
\Big] \tilde{\Psi}(\tilde{x},\tilde{y},t),
\end{array}
\end{equation}
where $\boldsymbol{\sigma}_i$ are the standard Pauli matrices acting
on the spinor space and
\begin{equation}
\mathbf{g}=\left[\begin{array}{cc} \tilde{g}_{11} & \tilde{g}_{12} \\
\tilde{g}_{12} & \tilde{g}_{22}\end{array}\right]
\end{equation}
is the matrix valued scattering amplitude. In general, the
scattering amplitudes $\tilde{g}_{11},\,\tilde{g}_{22}$ and
$\tilde{g}_{12}$ can be different. We will in the following assume
$\tilde{g}_{11}=\tilde{g}_{22}=\tilde{g}$ and $\tilde{g}_{12}=0$, a
condition that is supposed to be approximately achievable for a
$^{87}$Rb condensate \cite{gaugegen}. We have numerically verified,
using a non-zero $\tilde{g}_{12}$ does not give rise to any
qualitative changes of our results. In terms of physical quantities,
$\tilde{g}=4\pi\hbar^2Na/mV$, where $N$ is the number of condensate
atoms, $V$ the effective volume and $a$ the $s$-wave scattering
length. Attractive atom-atom interaction is obtained for
$\tilde{g}<0$ and repulsive for $\tilde{g}>0$. Noteworthy is that
the parameter $\tilde{g}$ may be tuned over a wide range of values
using Feshbach resonance techniques \cite{fb}.

The single particle Hamiltonian Eq. (\ref{ham0}), once the harmonic
trapping potential has been included, is equivalent to the
$E\times\varepsilon$ Jahn-Teller one, which was first introduced in
molecular physics \cite{lh}. However, the canonical momentum and
position variables are interchanged in the original
$E\times\varepsilon$ Hamiltonian. Mathematically this is just a
question of representations, but it has physical consequences, for
example in state preparation and measurement. Another outcome of this
is that the effect of the Berry phase normally apparent in the
physical state $\tilde{\Psi}(\tilde{x},\tilde{y},t)$ will instead
manifest itself in the spinor wave function
$\tilde{\Phi}(\tilde{p}_{\tilde{x}},\tilde{p}_{\tilde{y}},t)$ in the
momentum representation, $\tilde{\Phi}$ being related to
$\tilde{\Psi}$ via the Fourier transform
\begin{equation}
\tilde{\Phi}(\tilde{p}_{\tilde{x}},\tilde{p}_{\tilde{y}},t) =
\int d\tilde{x}d\tilde{y}\,e^{-i(\tilde{x}\tilde{p}_{\tilde{x}} +
\tilde{y}\tilde{p}_{\tilde{y}})/\hbar}
\tilde{\Psi}(\tilde{x},\tilde{y},t).
\end{equation}
The phenomena we are studying are therefore more conveniently
extracted by analyzing the problem in a momentum representation, and
in particular in terms of an effective momentum-dependent potential.

\begin{figure}
\centerline{\includegraphics[width=6.5cm]{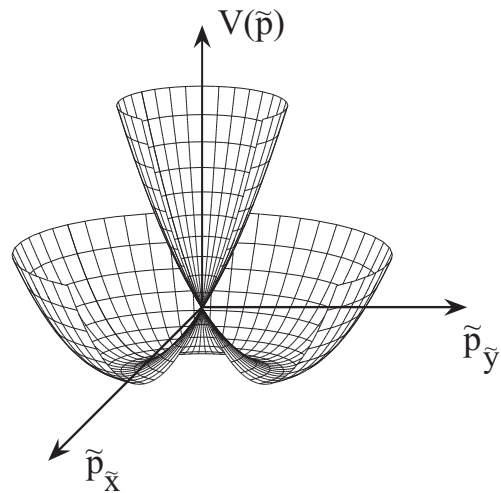}}
\caption{Effective APS corresponding to the Gross-Pitaevskii
equation (\ref{ham}). The conical intersection is located at
the origin, $\tilde{p}_{\tilde{x}}=\tilde{p}_{\tilde{y}}=0$,
where the two APS become degenerate. Here we have
$\tilde{v}_0=\tilde{v}_1$. }
\label{fig2}
\end{figure}

For this purpose we introduce the unitary adiabatic transformation
$U_{\mathrm{ad}}=U_{\mathrm{ad}}(\tilde{p}_{\tilde{x}},
\tilde{p}_{\tilde{y}})$ that diagonalizes the spin-orbit
term, namely
\begin{equation} \label{eq:at}
\begin{array}{l}
U_{\mathrm{ad}}\left[\begin{array}{cc} 0 &
\tilde{v}_0\tilde{p}_{\tilde{x}}-i\tilde{v}_1\tilde{p}_{\tilde{y}}\\
\tilde{v}_0\tilde{p}_{\tilde{x}}+i\tilde{v}_1\tilde{p}_{\tilde{y}} & 0
\end{array}\right]U_{\mathrm{ad}}^{\dagger}\\ \\
=\sqrt{\left(\tilde{v}_0\tilde{p}_{\tilde{x}}\right)^2 +
\left(\tilde{v}_1\tilde{p}_{\tilde{y}}\right)^2}
\boldsymbol{\sigma}_z \equiv \lambda \boldsymbol{\sigma}_z .
\end{array}
\end{equation}
This results in the effective adiabatic potential surfaces (APS) in
momentum space, defined as
\begin{equation}
V_{\mathrm{ad}}^\pm=\frac{1}{2m}\left(\tilde{p}_{\tilde{x}}^2+
\tilde{p}_{\tilde{y}}^2\right)\pm\lambda
\end{equation}
which conically intersect at $\tilde{p}=0$. Encircling the
conical intersection (CI) gives rise to a Berry phase \cite{berry},
which, indeed, is related to the gauge properties of the problem
\cite{molberry,ABmol}. Figure
\ref{fig2} displays the APS in the polar symmetric case in which
$\tilde{v}_0=\tilde{v}_1$. For $\tilde{v}_0\neq\tilde{v}_1$, on the
other hand, the polar symmetry is broken and the lowest APS possesses
two local minima instead of the symmetric Mexican hat sombrero
shape. In what follows, we will only consider the situation with
$\tilde{v}_0=\tilde{v}_1\equiv \tilde{v}$, a condition that can be met
by adjusting the laser parameters. Thus, in our simulations, the lower
APS in momentum space has the familiar sombrero structure.

\section{Dynamics}
\label{sec3}
We numerically solve the Gross-Pitaevskii equation (\ref{ham}) using
the split-operator method \cite{split}. First, we introduce
dimensionless variables scaled by the characteristic length and
energy given by $l=\sqrt{\hbar/m\omega}$ and $E_c=\hbar\omega$,
respectively. The Gross-Pitaevskii equation then attains the form
\begin{equation}\label{hamscal}
\begin{array}{lll}
i \frac{\partial}{\partial\tau}\Psi(x,y,\tau)\!\!\! & = \!\!\!\!&
\Bigg\{-\frac{1}{2}\left(\frac{\partial^2}{\partial
x^2}+\frac{\partial^2}{\partial y^2}\right) +
\frac{1}{2}\left(x^2+y^2\right) \\ \\
& & \!\!\!\!+ v \left[\begin{array}{cc} 0 & -i\frac{\partial}{\partial x}
-\frac{\partial}{\partial y}\\
-i\frac{\partial}{\partial x}+\frac{\partial}{\partial y} &
0\end{array}\right]
\\ \\
& & \!\!\!\!+\!g\!\left(\!|\psi_1(x,y,\tau\!)|^2 \!\! +
\!\!|\psi_2(x,y,\tau\!)|^2\!\right)\!\!\!\Bigg\}\!\!\Psi(x,y,\!\tau\!)\!,
\end{array}
\end{equation}
where
\begin{equation}
\begin{array}{lll}
x=\frac{\tilde{x}}{l}, & y=\frac{\tilde{y}}{l}, & \tau=t\omega, \\ \\
v=\frac{\tilde{v}}{E_c}, & g=\frac{\tilde{g}}{E_c}. &
\end{array}
\end{equation}
As initial spinor components, we take minimum uncertainty
Gaussians
\begin{eqnarray}
\psi_i(x,y,0) & = & a_i \left(\frac{1}{\pi\Delta_l^2}\right)^{1/2}
e^{-i(p_{x0}x+p_{y0}y)}
\nonumber \\
 & & \times e^{-\frac{(x-x_0)^2+(y-y_0)^2}{2\Delta_l^2}},
\end{eqnarray}
where $|a_i|^2$, normalized as $|a_1|^2+|a_2|^2=1$, determine the
initial population in states $|u_i\rangle$ ($i=1,2$), $\Delta_l$ is
the initial wave packet width and $x_0$, $y_0$, $p_{x0}$ and
$p_{y0}$ are initial averages of positions and momentum. We will
choose $\Delta_l=1$, corresponding to the ground state width of the
oscillator potential $\frac{1}{2}(x^2+y^2)$.

An advantage of utilizing the split-operator method is that the
time-dependent solution of the GP equation is automatically obtained
in both momentum and position representation. As already pointed
out, the effect of the Berry phase is most easily extracted
from the momentum wave packets. More precisely, the split-operator
method makes use of the fact that for sufficiently short time steps
$\delta\tau$, the time-evolution operator can be factorized into one
momentum-dependent and one spatially dependent part
\begin{equation}
\begin{array}{lll}
\Psi(x,y,\tau\!+\!\delta\tau)\!\! & =\!\! &
\displaystyle{\exp\!\left\{\!\!-i\!\!\left[\frac{1}{2}\!
\left(p_x^2\!+\!p_y^2\right)\!+\!v p_x\boldsymbol{\sigma}_x\!\!+
\!v p_y\boldsymbol{\sigma}_y\!\right]\! \delta\tau\!\right\}}\\
\\
& & \!\!\!\displaystyle{\times
\!\exp\!\left\{\!-i\!\!\left[\frac{1}{2}\!\left(x^2\! +
\!y^2\right)\!+g\!|\Psi(x,y,\tau)|^2\right]\!\delta\tau\!\right\}}\\
\\
& & \times\Psi(x,y,\tau),
\end{array}
\end{equation}
in the limit $\delta\tau\rightarrow0$. The first exponent multiplies
$\Psi(x,y,\tau)$, then the resulting wave packet is transformed to
momentum space via the fast Fourier algorithm and the second
exponent multiplies the momentum wave packet. Finally, the inverse
Fourier transform gives the propagated state $\Psi(x,y,\tau+\delta\tau)$.
The time step $\delta\tau$ is chosen such that contribution from the
non-commuting part, due to factorizing the time-evolution operator,
can be safely neglected.

\subsection{Berry phase over short time periods}
\label{ssec3a} 
\begin{figure}
\centerline{\includegraphics[width=8cm]{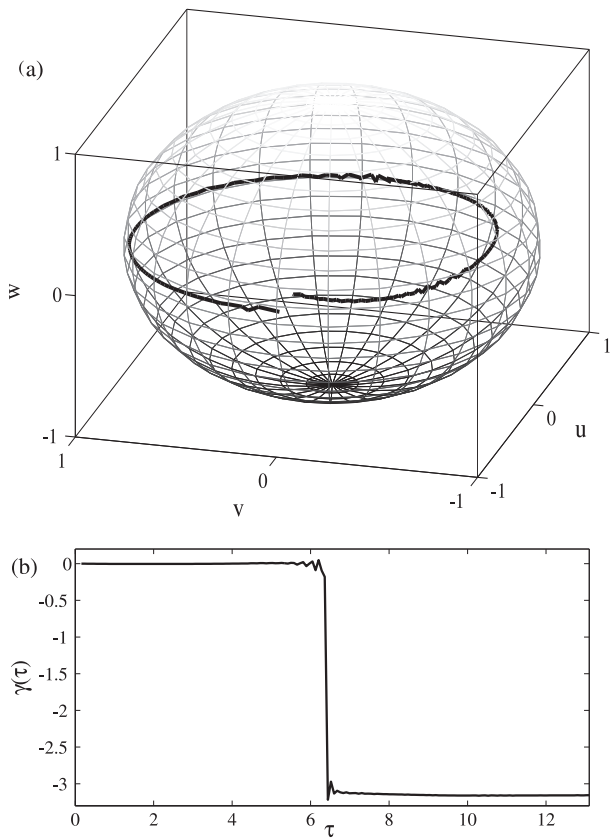}}
\caption{Bloch vector and corresponding non-cyclic Berry phase for
the spinor subsystem with no atom-atom scattering ($g=0$). The
initial state is characterized by the dimensionless parameters $a_1=-a_2=
1/\sqrt{2}$, $x_0=0$, $y_0=-3$, $p_{x0}=4$, and $p_{y0}=0$.}
\label{fig:figberry}
\end{figure}

Let us first examine how the Berry phase of the
spinor part of the state behaves when the wave packet encircles the
origin $(p_x,p_y)=(0,0)$. We mainly consider the case with no
atom-atom scattering ($g=0$), and only briefly discuss the $g\neq0$
case. The spinor state is given by the reduced density operator
\begin{eqnarray}
\rho (t) & = & P_1 (\tau) |u_1\rangle \langle u_1| +
P_2 (\tau) |u_2\rangle \langle u_2|
\nonumber \\
 & & + C (\tau) |u_1\rangle \langle u_2| +
C^{\ast} (\tau) |u_2\rangle \langle u_1| ,
\end{eqnarray}
where $P_i (\tau) = \langle \psi_i(\tau)|\psi_i(\tau)\rangle$,
$i=1,2$, and $C(\tau) = \langle \psi_1(\tau)|\psi_2(\tau)\rangle$
represent the populations and interference, respectively, of the two
dark states $u_1,u_2$. By writing $\rho(\tau)$ on Bloch form, we 
identify the time dependent Bloch vector ${\bf r}(\tau) = 
(u(\tau),v(\tau),w(\tau)) = (2{\textrm{Re}} C(\tau), -2{\textrm{Im}} 
C(\tau),P_1(\tau) - P_2 (\tau))$. We choose an initial wave packet 
characterized by the parameters
$a_1=-a_2=1/\sqrt{2}$, $x_0=0$, $y_0=-3$, $p_{x0}=v=4$, and
$p_{y0}=0$. For this choice, the wave packet will predominantly
populate the lower APS and in particular since $p_{x0}=v$ it is
located at the minima of the sombrero \cite{SK,jonasJT}. Moreover,
due to the non-zero $y_0 < 0$, as time evolves it traverses the sombrero
minima clockwise. The motion of the Bloch vector representing the
spinor state is shown in the upper panel of Fig.
(\ref{fig:figberry}). Note that $|{\bf r}(0)|=1$ (pure state) since
$|\psi_1 (0) \rangle = |\psi_2 (0) \rangle \equiv |\psi (0) \rangle
$, which implies that the initial spinor wave function takes the
product form $|\Psi (0)\rangle = |\psi (0) \rangle \left( |u_1\rangle -
|u_2 \rangle \right) /\sqrt{2}$. The Jahn-Teller coupling creates
entanglement between the spinor and spatial degrees of freedom when
the wave packet evolves, which explains why $|{\bf r} (\tau)|$
varies. To compute the evolving Berry phase $\gamma (\tau)$ for
this effective non-unitary evolution of the spinor subsystem we
write the reduced density operator on spectral form $\rho (\tau) =
\sum_{k=1}^2 \lambda_k (\tau) |\varphi_k (\tau) \rangle \langle
\varphi_k (\tau)|$ and use \cite{tong04}
\begin{eqnarray}\label{berryexp}
\gamma (\tau) & = &
\arg \left[ \sum_{k=1}^2 \sqrt{\lambda_k (0) \lambda_k (\tau)}
\langle \varphi_k (0)| \varphi_k (\tau) \rangle \right.
\nonumber \\
 & & \left. \times e^{-\int_0^{\tau} \langle \varphi_k (\tau') |
\dot{\varphi}_k (\tau') \rangle d\tau'}\right]
\nonumber \\
 & = & \arg \left[ \langle \varphi_1 (0)| \varphi_1 (\tau) \rangle \right]
\nonumber \\
 & & + i \int_0^{\tau} \langle \varphi_1 (\tau') |
\dot{\varphi}_1 (\tau') \rangle d\tau',
\end{eqnarray}
where the second equality follows from that $\rho (0)$ is pure (only
one $\lambda_k (0)$ is nonzero) and the assumption that $\lambda_1
(\tau)$ is the largest eigenvalue over the relevant time interval.
Thus, the Berry phase in this case becomes the standard non-cyclic
geometric phase \cite{samuel88} for the eigenstate of $\rho(\tau)$
with largest eigenvalue. Here, this phase becomes $-\frac{1}{2}$ times
the solid angle $\Omega_{\textrm{gc}} (\tau)$ enclosed by the path
$\tau' \mapsto {\bf n} (\tau') = {\bf r} (\tau')/|{\bf r} (\tau')|$
and the shortest geodesics that connects its end-points ${\bf n} (0)$
and ${\bf n} (\tau)$. From the upper panel of Fig.~\ref{fig:figberry}
we see that $\Omega_{\textrm{gc}} (\tau) \approx 0$ for the first half
of the path, and $\Omega_{\textrm{gc}} (\tau) \approx 2\pi$ for the
second half.  Thus, we expect $\gamma (\tau) = -\frac{1}{2}
\Omega_{\textrm{gc}} (\tau)$ to make a $\pi$-jump at half the period
of the motion. This behavior of the non-cyclic Berry phase in the
$E\otimes\varepsilon$ Jahn-Teller system has been examined in the
literature \cite{sjoqvist97} and is verified in the lower panel of
Fig.~\ref{fig:figberry}. Note the wiggles in the vicinity of the phase
jump which are caused by small non-adiabatic effects in the wave
packet motion.

To extract measurable effects of the non-zero Berry phase we need
the wave packet to self-interfere. To this end, we assume as above
$a_1=-a_2=1/\sqrt{2}$, $x_0=y_0=p_{y0}=0$, $p_{x0}=v$, while now
$y_0=0$. Thus, the momentum wave packet has no initial velocity, so
for zero atom-atom scattering corresponding to $g=0$, the wave
packet starts to spread around the minima of the sombrero. We
emphasize that this spreading takes place in the momentum wave
packet $\Phi$, and is driven by the uncertainty in the spatial
coordinates. Thus, such spreading is only possible since we consider
a harmonically trapped condensate, contrary to
Refs.~\cite{berryCA1,rev2} where no trapping was taken into account.
For positive $g$, the interaction is repulsive which tends to
increase the spreading of the $\Psi$ wave packet while slowing down
the $\Phi$ wave packet broadening. The opposite occurs for
attractive interaction. Consequently, no overall spreading in $\Phi$
takes place if $g$ is large and negative.

We assume a $g$ such that for sufficiently long times, $\Phi$ has
spread noticeable compared to its initial state. Then, after some
time $\tau_{\mathrm{col}}$, the wave packet has expanded a distance
$2\pi\rho_{\mathrm{min}}$, where $\rho_{\mathrm{min}}$ is the radii
of the minima of the sombrero. The wave packet tails will then start
to overlap and cause a self-interference pattern. It was shown in
Ref.~\cite{jonasJT} that $\tau_{\mathrm{col}}=2\pi v$, a result
which applies to the present model in the absence of atom-atom
interaction. As the structure of the interference depends on the
wave packet phases, both the dynamical phase and the Berry phase
will be manifested in its shape.

\begin{figure}[h]
\centerline{\includegraphics[width=8cm]{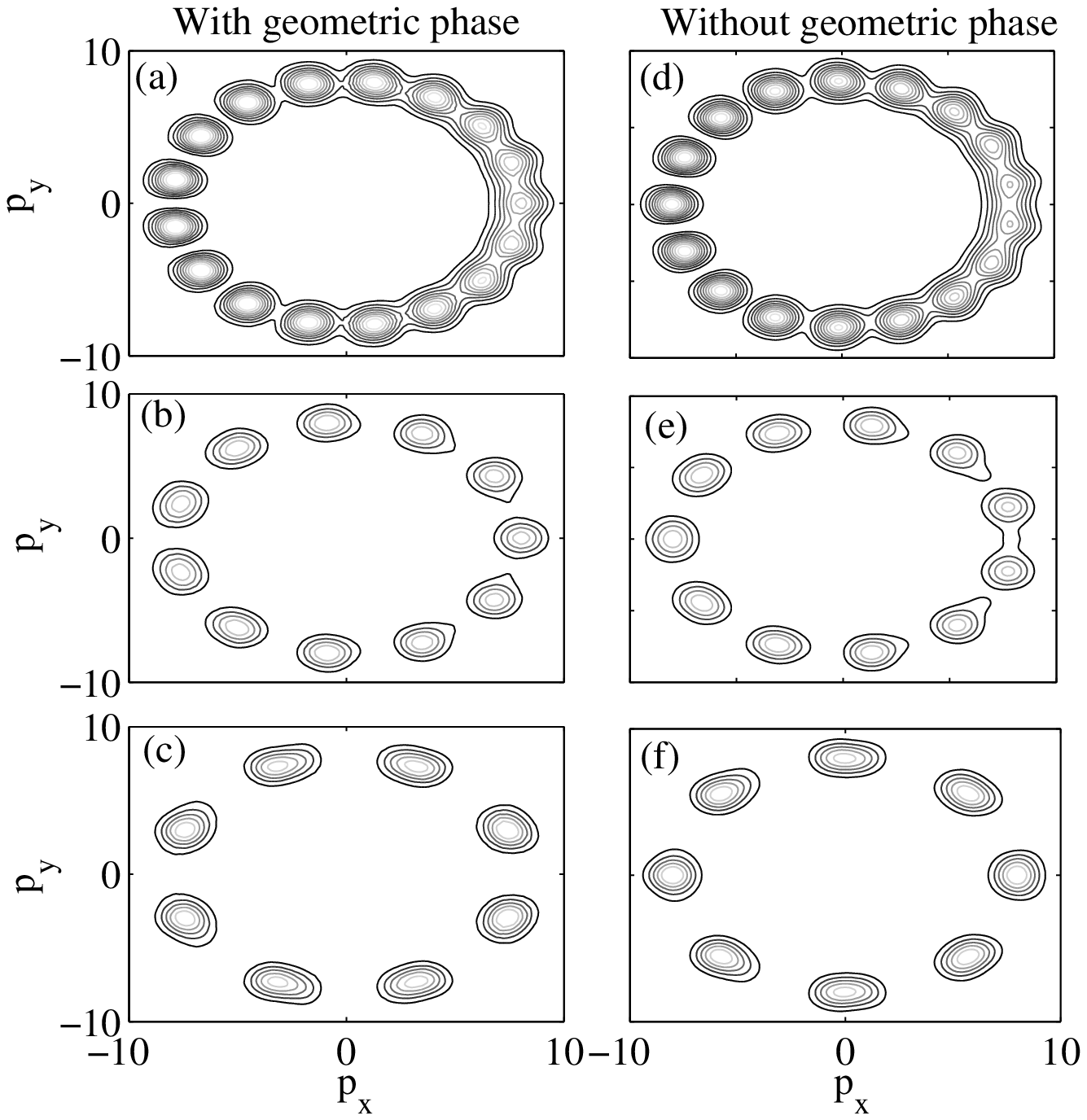}}
\centerline{\includegraphics[width=8cm]{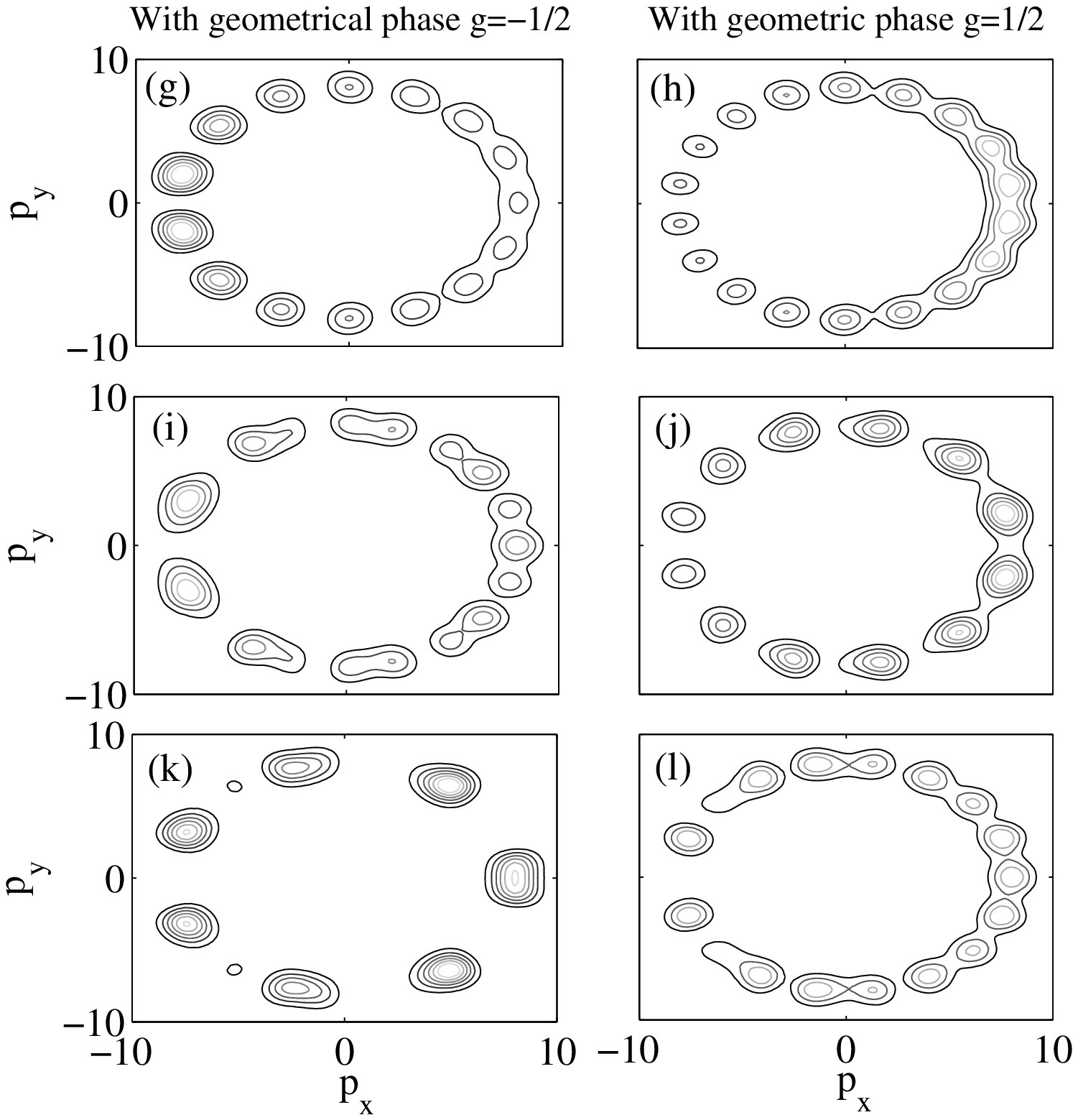}}
\caption{Snapshots at times $\tau = \tau_{\mathrm{col}}/2,\,3
\tau_{\mathrm{col}}/4, \, \tau_{\mathrm{col}}$ of the momentum
distribution Eq. (\ref{momdist}) after the initial wave packet has
spread out around the minima of the sombrero potential. Upper left
plot presents the result with a Berry phase present and upper right
one without it. Here, the dimensionless parameters are $v=8$, $g=0$,
and $\tau_{\mathrm{col}}=16\pi$. The lower plots (g)-(l) display the
results from having a non-zero scattering amplitude ($g\neq 0$),
essentially affecting the rate of wave packet broadening. Again,
$v=8$ and $\tau_{\mathrm{col}}=16\pi$.}
\label{fig3}
\end{figure}

In order to describe the effect of the Berry phase $\gamma$, the
dynamics rendered by the Gross-Pitaevskii equation (\ref{hamscal})
will be compared with the dynamics of an adiabatic Hamiltonian
possessing the same lower APS but lacking any Berry phase. As
mentioned above, for the particular choice of parameters,
$x_0=y_0=p_{y0}=0$, $p_{x0}=v$, $\Delta_l=1$ and
$a_1=-a_2=1/\sqrt{2}$, the wave packet evolution takes place
essentially on the lower APS. We therefore make a {\it single
surface approximation} \cite{ssa}, writing an effective Hamiltonian
as
\begin{equation}\label{ssham}
H_{\mathrm{S}}=\frac{x^2}{2}+\frac{y^2}{2}+V_{\mathrm{ad}}^- ,
\end{equation}
which follows from the adiabatic transformation defined in Eq.
(\ref{eq:at}) and by ignoring the Berry vector potential
$\vec{A} (\vec{p})= (-p_y,p_x)/(2\vec{p}\,^2)$ as well
as the Born-Huang scalar potential $1/(8\vec{p}\,^2)$
\cite{born54,berry90}. Ignoring $\vec{A}$ implies that no
Berry phase arises when the wave packet encircles the CI at the origin
in momentum space \cite{jonasJT}. On the other hand, the dynamical
phases acquired in the two models are almost identical, since the
effect of the Born-Huang potential is small for the type of evolution
that is considered here. In order to identify the consequences of the
Berry phase on the characteristics of the two systems, it is therefore
convenient to analyze the wave packet evolution in the two respective
models. Figure \ref{fig3} depicts, for $g=0$ (a)-(f) and
$g\neq0$ (g)-(l), the momentum distribution
\begin{equation}\label{momdist}
|\Phi(p_x,p_y,\tau_f)|^2 = |\phi_1(p_x,p_y,\tau_f)|^2 +
|\phi_2(p_x,p_y,\tau_f)|^2
\end{equation}
from our numerical propagation of the wave packet after a time
$\tau_f=\tau_{\mathrm{col}}$ when self-interference has been clearly
manifested. Similar figures were presented in Refs.~\cite{SK,jonasJT}
for different systems. The upper left and lower plots are obtained
from the full system evolution governed by Eq.~(\ref{hamscal}), while
the upper right plots are the results achieved using the single
surface approximation (\ref{ssham}). The effect of the Berry phase is
evident; it is especially found that a node/antinode appears at
$p_x<0,\,p_y=0$ in the cases with/without Berry phase. The location of
this node/antinode persists throughout the evolution. The total number
of nodes/antinodes, on the other hand, changes with time, which was
demonstrated in Ref.~\cite{jonasJT}.

The wave packet spreading, as it evolves along the minima of the
sombrero, takes place relatively far from the CI for the parameters
of the above examples. In other words, the dynamics is predominantly
adiabatic. However, the concept of adiabaticity in non-linear
systems is highly non-trivial since the superposition principle and
orthonormality of quantum states are no longer applicable
\cite{niu}. As the strength of the non-linearity is increased, the
constraints for adiabatic evolution are in general strengthened.
This has been studied in numerous papers, mainly focusing on various
curve-crossing models \cite{nlad,loop}, but also on the Berry phase
\cite{nlberry}. In the current system and for the parameters
considered, non-linearity manifests itself in the characteristic
time scales rather than on the Berry phase effect. This is
illustrated in Fig.~\ref{fig3} (g)-(l) where the Berry phase induced
node at $p_x<0,\,p_y=0$ is invariant. The number of nodes, however,
depends on the non-linearity coefficient $g$, and consequently the
characteristic times are $g$-dependent. This result may be taken as
further evidence of the conjectured robustness of Berry phase
effects to various kinds of errors and therefore of the potential
utility of such phases in the implementation of robust quantum
computers \cite{jones00}. The persistence of the Berry phase effect
due to non-linearity was also demonstrated in the model of
Ref.~\cite{berrybec2}. We should point out that in other situations
than the one considered in this work, namely when the wave packet
dynamics takes place at the CI, the effect of the non-linearity is
supposed to become important. It is known that non-linearity of the
periodic Gross-Pitaevskii equation can bring about loops at the
Brillouin center and at its edges \cite{loop}, and similar loops
might be formed at the CI. This, however, goes beyond the scope of
this work.

\begin{figure}[h]
\centerline{\includegraphics[width=8cm]{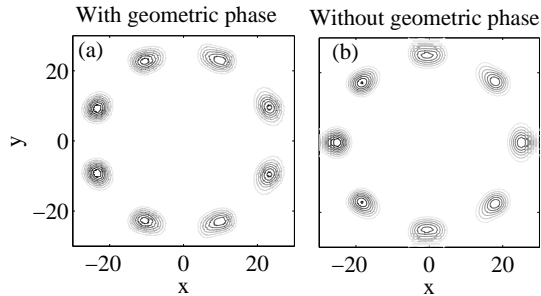}} \vspace{-1.5cm}
\caption{The position distributions after ballistic expansion. As in
Fig.~\ref{fig3}, the left and right plots represent the results with
and without a Berry phase. The dimensionless parameters are the same as
in Fig.~\ref{fig3} (a)-(f), and the time-of-flight duration is
$\tau_{\textrm{tof}}=\pi$.} \label{fig4}
\end{figure}

To experimentally extract the impact of the Berry phase, a
time-of-flight measurement can be implemented by switching off the
trapping potential and the external lasers to let the condensate
ballistically evolve before detecting it. In Fig.~\ref{fig4} we show
an example of such a ballistic expansion of the wave packet
distributions $|\Psi(x,y,\tau)|^2$ after it has been released from the
trap. The parameters are the same as in Fig.~\ref{fig3} (a)-(f) and the
release time $\tau_r=\tau_{\mathrm{col}} = 16\pi$ corresponds to
Fig.~\ref{fig3} (c) and (f). The form of the momentum distributions,
as shown in Fig.~\ref{fig3} (c) and (f), are clearly embodied in the
position distribution. Assuming a typical trapping potential frequency
$\omega/2\pi=40$ Hz, the characteristic time scale is 4 ms. Hence, the
time-of-flight of Fig.~\ref{fig4} is around 10 ms.  This is of the
same order as in common time-of-flight measurements of condensates in
optical lattices \cite{mott}. Moreover, the interaction times of
Fig.~\ref{fig3} are of the order of 50-100 ms for this particular
trapping potential frequency. Quantum phase diffusion of the
condensate \cite{qfd}, arising from quantum fluctuations beyond the
mean field description, will most likely decrease the visibility of
the interference pattern. However, phase coherence of 50 ms has been
observed experimentally \cite{qfd2}, and at these time scales the
Berry phase has already been manifested in the interference pattern.

\subsection{Impact of the Berry phase over long times}\label{ssec3b}

\begin{figure}[h]
\centerline{\includegraphics[width=8cm]{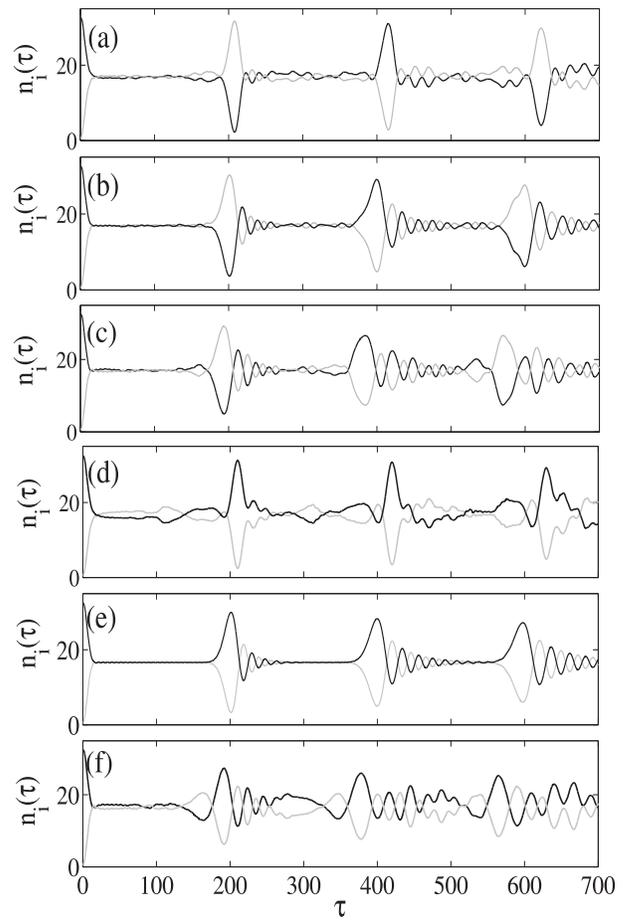}}
\caption{The average number of phonon excitations $n_i$ $i=x$ (black
curve) and $i=y$ (gray curve) as a function of time with (a)-(c) and
without (d)-(f) Berry phase. Here $g=0.25$ (a) and (d), $g=0$ (b)
and (e) and $g=-0.25$ (c) and (f). The rest of the dimensionless
parameters are as in Fig.~\ref{fig3}. } \label{fig5}
\end{figure}

The time scale $\tau_{\mathrm{col}}$ for self-interference of the wave
packet to be established, characterizes a collapse period. For longer
time periods, the number of wave packet maxima/minima of Fig.~\ref{fig3}
decrease to finally a single maximum is recovered signaling a full or
half revival \cite{jonasJT,robinett}. The formation of the wave packet
maxima/minima is indeed highly dependent on the Berry phase. To verify
this, we define the number of phonon excitations in the two modes
\begin{equation}
n_i\equiv\frac{\langle p_i^2\rangle}{2} +
\frac{\langle i^2\rangle}{2},\hspace{1.2cm}i=x,y.
\end{equation}
This quantity is depicted in Fig.~\ref{fig5}, where the upper three
plots are found using the full Hamiltonian Eq. (\ref{hamscal}) and the
corresponding single surface results are presented in the lower three
plots. In (a) and (d) $g=0.25$, in (b) and (e) $g=0$ and finally in
(c) and (f) $g=-0.25$. Interestingly, at around $\tau\approx200$ the
population is swapped between the two phonon modes in the case of
Berry phase (a)-(c) but not when the Berry phase is absent
(d)-(f). At times $\tau\approx400$ does a half revival occur
\cite{jonasJT}. The figure also gives a measure on how the non-linear
interaction affects the time scales: positive $g$ increases while
negative $g$ decreases the time periods over which the system
characteristics are established.

\subsection{Non-Abelian manifestation}\label{ssec3c}
\begin{figure}[h]
\centerline{\includegraphics[width=8cm]{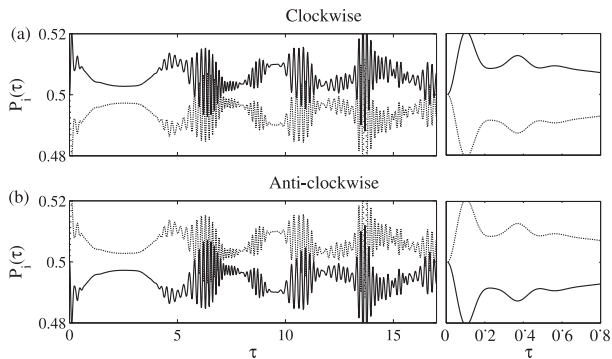}}
\caption{Populations $P_i(\tau)$ of the dark states $|u_i\rangle$
for clockwise and anti-clockwise evolution of the wave packet around
the CI. Solid lines show the population of the state $|u_2\rangle$
and dashed lines the population of $|u_1\rangle$. The figures to the
right display the short time behavior, which evidence that initially
both internal states are equally populated $P_1(0)=P_2(0)=1/2$. The
initial state are as in the previous figures ($p_{x0}=v$,
$\Delta_l=1$ and $a_1=-a_2=1/\sqrt{2}$) and $y_0=3$ (a) or $y_0=-3$
(b). The other dimensionless parameters are $g=0$ and $v=4$.}
\label{fig6}
\end{figure}

We note that the Hamiltonian of Eq.~(\ref{hamscal}) can be written,
for $g=0$, as
\begin{equation}
H_{\mathrm{eff}}=\frac{\left(\vec{p}-\vec{A}\right)^2}{2} +
\frac{\left(x^2+y^2\right)}{2},
\end{equation}
where $\vec{p}=(p_x,p_y)$ and $\vec{A} = (A_x,A_y) =
v(\boldsymbol{\sigma}_x,\boldsymbol{\sigma}_y)$. The gauge vector
potential $\vec{A}$ has a matrix form due to the internal two-level
structure of our system. The fact that $[A_x,A_y] \neq0$ implies
that the gauge field has a non-Abelian character. As pointed out in
Ref.~\cite{berryCA1}, encircling the CI clockwise or anti-clockwise
is supposed to bring about different system evolution despite the
polar symmetry of the system. More precisely, after one round-trip
of the CI, the populations $P_i(\tau)$ of the two degenerate dark
states $|u_i\rangle$ ($i=1,2$) depend on the direction the wave
packet traversed the minima of the sombrero. We demonstrate this in
Fig.~\ref{fig6} by displaying the time evolution of the populations
for approximately one round-trip clockwise (a) or anti-clockwise
(b). We find that the progression of $P_1(\tau)$ and $P_2(\tau)$ is
interchanged between clockwise or anti-clockwise wave packet
evolution, and therefore revealing the underlying non-Abelian
structure. From the right plots, zooming in on the short time
evolution, it is clear that the initial population is equally
balanced between the two internal states, i.e.,
$|a_1|^2=|a_2|^2=1/2$.

\section{Conclusions}\label{sec4}
We have analyzed a spin-orbit coupled BEC system, emphasizing on the
Berry phase and non-Abelian gauge structures and their importance for
the dynamics.  We have demonstrated that the non-Abelian structure
associated with the center of mass position of the atoms is mirrored
in the momentum space wave packet interference as a Berry phase. This
Berry phase has been computed and its effect has been identified by
comparing the dynamics of the full system with one possessing the same
adiabatic potential energy surface but lacking the Berry phase. A
time-of-flight measurement would therefore directly bring out the
presence of such a phase. Over longer time periods, a periodic Berry
phase dependent interchange of excitations between the $x$ and $y$
directions was demonstrated. We have also shown how the non-Abelian
aspect of the model system is manifested in the wave packet dynamics
in position space by propagating in different directions around the
CI.

It is worth mentioning that the results are not restricted to a
spin-orbit coupled BEC, but can equally well be applied to single
trapped ions \cite{iontrap}. Using the same laser and atomic
configurations, the dynamics is governed by Eq. (\ref{hamscal})
letting $g\rightarrow0$. Berry phases have been discussed in such
systems \cite{berryion}, but those settings are different from the
one considered here.

\begin{acknowledgments}
JL acknowledge support from the MEC program (FIS2005-04627) and
wishes to thank Prof. Stig Stenholm for fruitful discussions.
ES acknowledge support from the Swedish Research Council.
\end{acknowledgments}

\end{document}